\documentclass{PoS}
\usepackage{epsfig}

\PoS{PoS(LAT2005)119}

\title{Isospin breaking and the chiral condensate }

\ShortTitle{Isospin breaking and the chiral condensate }

\author{\speaker{Michael Creutz}\\
        Brookhaven National Laboratory\\
        E-mail: \email{creutz@bnl.gov}}

\abstract{With two degenerate quarks, the chiral condensate exhibits a
jump as the quark masses pass through zero.  I discuss how this single
transition splits into two Ising like transitions when the quarks are
made non-degenerate.  The order parameter is the expectation of the
neutral pion field.  The transitions represent long distance coherent
phenomena occuring without the Dirac operator having vanishingly small
eigenvalues.}

\FullConference{XXIIIrd International Symposium on Lattice Field Theory\\
		 25-30 July 2005\\
		 Trinity College, Dublin, Ireland}

\begin{document}

\begin{figure}
\centering
\includegraphics[width=3in]{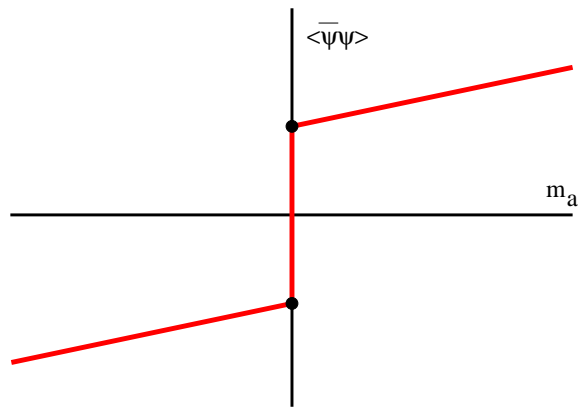}
\caption{
A sketch of how the chiral condensate jumps in sign as the quark mass
passes through zero for two flavors of degenerate quark.  }
\label{condensate0}
\end{figure}

Spontaneous chiral symmetry breaking is a cornerstone of our
understanding of low energy hadronic dynamics.  In particular, the
idea of pions as approximate Goldstone bosons is the accepted
explanation of why they are so much lighter than the rho mesons,
despite having the same quark content.  In this picture, a condensate
appears with $\overline\psi\psi$ maintaining an expectation value even
for vanishing quark mass.  The value of this condensate depends on the
direction from which the masses approach the chiral limit.  With two
flavors, one expects a jump in the chiral condensate as the quark mass
passes through zero.  This behavior is sketched in
Fig.~(\ref{condensate0}).  As a flavored chiral rotation can flip the
sign of the mass, the physics is the equivalent on both sides of the
transition.

\begin{figure}
\centering
\includegraphics[width=4in]{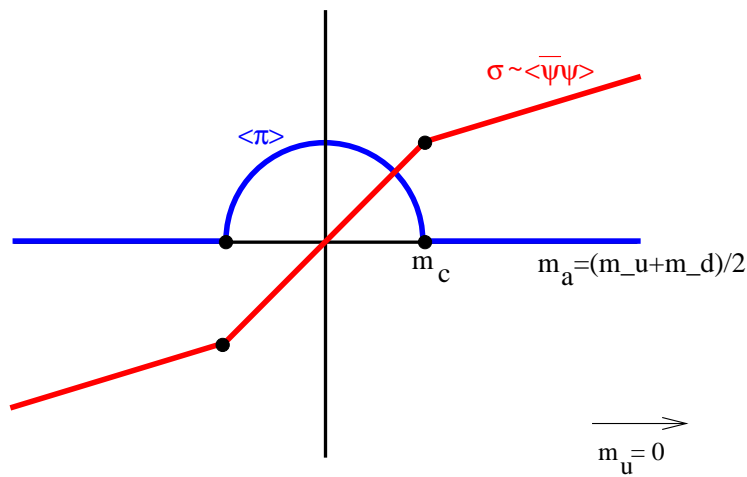}
\caption{ With a constant up-down quark mass difference, the jump in
the chiral condensate splits into two second order transitions.  The
order parameter distinguishing the intermediate phase is the
expectation value of the neutral pion field.  }
\label{condensate1}
\end{figure}

This represents the long established picture with two degenerate
quarks.  The question I address here is what happens when the quarks
are not degenerate.  In particular, hold the up-down quark mass
difference fixed and vary the average quark mass through zero.  The
result is a splitting of the transition into two second order
transitions, as sketched in Fig.~(\ref{condensate1}).  These
transitions are each Ising like and the order parameter is the
expectation value of the neutral pion field, $\langle \pi_0 \rangle$.
The effect is quadratic in the splitting of the quark masses, the
critical value for the average quark mass $m_c$ being proportional to
$(m_d-m_u)^2$.  Since the neutral pion is odd under CP, this symmetry
is spontaneously broken in the intermediate phase.  Finally I note
that these transitions have nothing to do with the point where the up
quark mass vanishes.  Indeed, the latter point is unphysical since it
depends on the detailed regularization scheme \cite{Creutz:2004fi}.

An effective chiral Lagrangian provides the simplest approach to see
how this structure arises.  I build on the simplest linear sigma model
involving the fields $\sigma$ and $\vec\pi$ interacting with a
potential
$$
V=\lambda(\sigma^2+\vec\pi^2-v^2)^2-m\sigma
$$ Here $v$ represents the vacuum expectation value of $\sigma$ and
$m$ is proportional to the average quark mass.  The four meson
coupling is denoted $\lambda$.  The $\sigma$ field models the
condensing combination $\overline\psi \psi$.  The chiral transition
occurs at $m=0$, where the order parameter $\langle\sigma\rangle$
jumps between $v$ and $-v$.

I now introduce isospin breaking into the picture.  This will come
from a non-vanishing mass difference between the down and up quarks.
Without isospin symmetry, there is no reason for the neutral and
charged pions to be degenerate.  In particular, the breaking will
allow the neutral pion to mix with isoscalar mesons.  There are
several candidates for this mixing, including the eta, the eta prime,
and glueballs.  As expected in quantum mechanics, this mixing will
result in level repulsion.  As the neutral pion is the lightest
neutral state, the mixing inherently destabilizes the $\pi_0$ field.

Although several states are involved in this mixing, for pedagogy I
simplify things and consider only a single effective isoscalar field
$\phi$.  Adding such a meson to the effective potential suggests one
study
\begin{equation}
V=\lambda(\sigma^2+\vec\pi^2-v^2)^2-m\sigma
{-\xi \pi_0\phi+m_\phi^2 \phi^2/2}
\end{equation}
Here I have given the new field an initial mass $m_\phi$, and
introduced the parameter $\xi$ to control the mixing between $\phi$
and $\pi_0$.  The strength of this mixing should be proportional to
the quark mass difference, $\xi\propto m_d-m_u$.  Note that the sign
of the mixing parameter $\xi$ is irrelevant as it can be changed by
redefining $\phi$ as its negative.

Now consider minimizing the potential $V$ over fields.  This gives the
set of equations
\begin{eqnarray}
0&=&{\partial V\over \partial \sigma}
=4\lambda\sigma(\sigma^2+{\pi_0}^2-v^2)-m\cr
\cr
0&=&{\partial V\over \partial \pi_0}
=4\lambda\pi_0(\sigma^2+{\pi_0}^2-v^2)-\xi \phi\cr
\cr
0&=&{\partial V\over \partial \phi}
=m_\phi^2\phi-\xi \pi_0
\end{eqnarray}
The last of these allows elimination of $\phi$, resulting in 
\begin{eqnarray}
&\phi&={\xi\pi_0\over m_\phi^2}\cr
&\pi_0&\left(\sigma^2+\pi_0^2-v^2-{\xi^2\over 4\lambda m_\phi^2}\right)=0
\end{eqnarray}

There are two generic possibilities for the value of $\pi_0$.
At large $|m|$ the solution is CP conserving with 
\begin{eqnarray}
&\pi_0&\ =\ 0\cr
&\sigma&\ =\ \pm v\sqrt{1+{m\over 4 \lambda\sigma v^2}}\ \sim
\ \pm v.
\end{eqnarray}
However at small $|m|$ there is a lower energy and CP violating
solution
\begin{eqnarray}
&\pi_0&
=\pm v\sqrt{1-{m^2m_\phi^4\over v^2\xi^4}
+{\xi^2 \over 4\lambda v^2 m_\phi^2}}\cr
&\sigma&={m m_\phi^2\over \xi^2}
\end{eqnarray}
This possibility of a CP violating solution was noted some time ago by
Dashen \cite{Dashen:1970et}.  The transition between the two behaviors
occurs at the critical average mass
\begin{equation}
m_c=\pm {\xi ^2 v\over
m_\phi^2}\sqrt{1+{\xi^2\over 4\lambda v^2 m_\phi^2}}
\end{equation}
This is of quadratic order in the mixing parameter, and thus is
quadratic in the mass difference $m_d-m_u$.  The transition is
continuous, and the $Z_2$ symmetry of $\pi_0\leftrightarrow -\pi_0$
becomes spontaneously broken.  Thus we expect the transition to be
Ising like.

In these transitions the condensate difference $\langle\overline\psi_u
\psi_u-\overline\psi_d \psi_d\rangle$ is automatically driven to a
non-zero value by the explicit isospin breaking.  Indeed, this
difference should only display a rather weak dependence on average
quark mass.

There are several minor variations on this argument that give rise to
the same picture of a CP violating phase.  In
Ref.~\cite{Creutz:2003xu}, the three flavor theory was studied
containing a heavier strange quark; here mixing with the eta meson
drives the phenomenon.  A pure two flavor discussion
\cite{Creutz:1995wf} mixing the fields $\{\sigma,\vec\pi\}$ with the
combination $\{\eta,\vec\delta\}$, where $\vec\delta$ is an isovector
scalar field, showed a similar behavior.

As mentioned above, we expect this effect to be quadratic in the quark
mass difference.  The quark masses themselves should be small for or
the chiral Lagrangian approach to be valid.  Thus the critical mass
should be even smaller than the individual quark masses themselves,
{\it i.e.} $m_c=O(m_q^2) << |m_q|$.  Since the CP violation occurs
when the average quark mass is less than the individual quark masses,
the phenomenon requires quark masses of opposite sign.  Although
naively the signs of the quark masses can be changed by chiral
rotations, their relative sign is protected by the anomaly.
Furthermore, in this regime the sign of the fermion determinant is not
always positive.  This invalidates the Vafa-Witten \cite{Vafa:1984xg}
discussion suggesting that spontaneous CP violation cannot occur in
the strong interactions.

At the critical average quark mass, the neutral pion is massless.
This occurs at a point where neither of the quarks is itself massless.
This shows that it is possible to have a massless meson without
massless quarks.  Indeed, immediately around the point where a quark
mass vanishes, physics is analytic, as discussed by Di Vecchia and
Veneziano \cite{DiVecchia:1980ve}.  Because of this smooth behavior,
redefinitions of the underlying parameters result in an inherent
ambiguity in defining where, say, the up quark mass vanishes.  As
emphasized in Ref.~\cite{Creutz:2004fi}, this means that contrary to
popular lore, a vanishing up quark mass cannot be a fundamental
solution to the strong CP problem.

While it might be interesting to study these transitions numerically
in lattice gauge simulations, there are some technical obstacles that
must be faced.  As the fermion determinant is not positive when this
phenomenon occurs, simulations will be afflicted with a sign problem.
Furthermore, the mixing with isosinglet states occurs through
disconnected quark diagrams, which therefore must be included.

With a lattice regulator that violates chiral symmetry, this CP
violating phase can potentially survive even in the degenerate quark
limit.  This case represents the phase proposed some time ago by Aoki
\cite{Aoki:1986ua}, and explored in the context of effective
Lagrangians in Refs.~\cite{Creutz:1996bg,Sharpe:1998xm}.  Depending on
the sign of a certain parameter in the effective Lagrangian, the CP
violating phase can either survive in the degenerate mass case or it
could collapse into a single first order transition.

\begin{figure}
\centering
\includegraphics[width=3in]{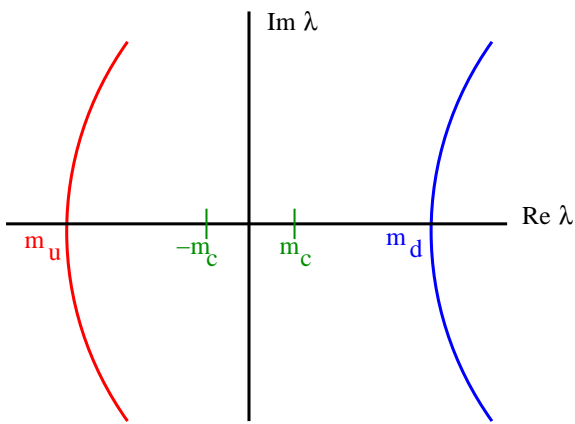}
\caption{ The eigenvalues of a Ginsparg-Wilson lattice fermion
operator are constrained to lie on large circles.  With a fixed
up-down quark mass difference, the point where the neutral pion mass
vanishes does not occur when either of these circles passes through
the origin.  Thus we can have long distance physics when none of the
eigenvalues are small.  }
\label{eigenvalues}
\end{figure}

It is sometimes argued that long distance physics is controlled by low
eigenvalues of the Dirac operator.  The phenomenon discussed here is a
counterexample to that concept.  Consider a formulation in terms of a
lattice Dirac operator satisfying the Ginsparg-Wilson
\cite{Ginsparg:1981bj} relation.  An example of such an operator is
provided by the overlap approach \cite{Neuberger:1997fp}.  These
operators force their eigenvalues onto large circles, the radii of
which go to infinity as the continuum is approached.  This is sketched
in Fig.~(\ref{eigenvalues}).  These circles pass through the real axis
at the corresponding bare quark masses.  However, as discussed above,
the transitions discussed in this paper do not occur when either quark
mass vanishes, but at an average quark mass tuned to a value much less
than the magnitude of either quark mass.  At these transitions long
distance physics manifests itself in the vanishing neutral pion mass,
but this occurs away from any vanishingly small eigenvalues of the
quark operators.

\section*{Acknowledgements}
This manuscript has been authored under contract number
DE-AC02-98CH10886 with the U.S.~Department of Energy.  Accordingly,
the U.S. Government retains a non-exclusive, royalty-free license to
publish or reproduce the published form of this contribution, or allow
others to do so, for U.S.~Government purposes.

\end{document}